# A study of the U.S. domestic air transportation network: Temporal evolution of network topology and robustness from 2001 to 2016

Leonidas Siozos-Rousoulis[a], Dimitri Robert[a], and Wouter Verbeke[a]

[a] Vrije Universiteit Brussel (VUB), Pleinlaan 2, 1050 Brussels, Belgium.

## Abstract

The U.S. air transportation network (ATN) is critical to the mobility and the functioning of the United States. It is thus necessary to ensure that it is well-connected, efficient, and robust. Despite extensive research on its topology, the temporal evolution of the network's robustness and tolerance remains largely unexplored. In the present paper, a temporal study of the domestic U.S. ATN was performed based on annual flight data from 1996 to 2016 and network analytics were used to examine the effects of restructuring that followed the 9/11 events along with the current state of the system. Centrality measures were computed to assess the system's topology and its global robustness. A node deletion method was applied to assess the network's tolerance by simulating a targeted attack scenario.

The study showed that the 9/11 terrorist attacks triggered vast restructuring of the network, in terms of efficiency and security. Air traffic expanded, as new airports and air routes were introduced. Airlines reconsidered their strategy and optimized their operations, thus allowing the network to recover rapidly and become even more efficient. Security concerns resulted in significant improvement of the network's robustness. Since 2001, the global traffic and topological properties of the U.S. ATN have displayed continuous growth, due to the network's expansion. On the other hand, the robustness of the system has not shown an improving tendency. Findings suggest that although the system's ability to sustain its operational level under extreme circumstances has lately improved, its tolerance to targeted attacks has deteriorated. The presented methodology may be applied on different network levels or different transportation networks, to provide a general perspective of the system's vulnerabilities.

# 1 Introduction

The U.S. airline industry has evolved significantly since the airline deregulation act of 1978, as is evidenced by the spectacular growth of commercial aviation in the United States during recent years. However, the current state of U.S. commercial aviation has been largely shaped by the 9/11 attacks and the drastic restructuring that followed on many scales. Main actors of the restructuring were the domestic airlines and the U.S. government. Major carriers reevaluated their business model (Franke, 2004; Ghobrial and Irvin, 2004) to retain a strategic advantage over emerging low-cost carriers (Tam and Hansman, 2003). For the U.S. government, point of focus was the protection of the U.S. air transportation network (ATN) from future disruptive events (Seidenstat, 2004) and particularly to terrorist attacks. Although such occurrences are generally rare, they can have crippling effects.

The U.S. ATN, like most air transportation networks, is a scale-free network and follows a power-law degree distribution. It is thus exceptionally vulnerable to intentional targeted attacks (Crucitti et al., 2003), since its properties deteriorate dramatically when critical nodes are isolated. Moreover, air transport systems are dynamic and inherently complicated. Therefore, the design of an efficient and robust ATN requires a multi-dimensional and dynamic research approach., which can benefit from a network analysis to identify critical infrastructure (Reggiani, 2013) and analyze properties that determine its dynamic behavior and robustness (Lordan et al, 2014a).

Topic of the present paper is a temporal study of ATN evolution, using complex network theory. The domestic U.S. ATN is chosen as an ideal real-life test case and is characterized by a high share of domestic flights (almost 88%) (Bureau of Transportation Statistics, 2017). Intention of the authors is to identify trends in the evolution of the network's topology and robustness, during the radical restructuring that occurred since 2001. Global indicators are used to analyse the network at a high level, which is important from a national security and an air traffic management perspective.

Annual flight data from 1996 till 2016 are analyzed to assess the impact of restructuring on its robustness. Robustness is defined as the combination of resilience and tolerance to targeted attack, thus being directly related to its topological properties and its connectivity. Initially, topological properties are assessed and centrality measures are computed. Global resilience of the system to disruptive events is calculated based on a parameter suggested by Janić (2015). The robustness study relies on a topology analysis, which allows to locate critical nodes and to detect scale-free properties that render the network vulnerable (Reggiani, 2013). Critical nodes are isolated and removed, thus simulating the worst-case scenario of a targeted attack (Chi and Cai, 2004). The network's efficiency and largest connected component are also monitored. The network's present state is finally analysed and managerial implications concerning its operations are discussed.

The present study adopts an unweighted high-level analysis. The methodology itself can be however easily replicated and applied on different levels of the U.S. ATN (e.g. airline or seasonal level), as well as on different transportation networks. Weights can be also accounted for (e.g. passenger capacity, seat pricing, or cargo value) depending on the different regions or transportation modes.

The structure of the paper is as follows: In section 2, the current scientific state of art is surveyed, whereas section 3 comprises of the mathematical and theoretical background of the research. The results of the study are presented and discussed in section 4. Finally, the conclusions of the paper are outlined in section 5.

# 2 Literature review

Air transportation networks are critical for the mobility of people. During the past few decades, the air transportation system has evolved into a complex and heterogeneous network (Barrat et al., 2004), having been shaped by geo-spatial relations, political constraints, and economical developments, among others (Guimerà et al., 2005; Wuellner et al., 2010). Hence, the dynamics of air transport networks have been lately the research focus in network analysis, complexity science, and geography (Vowles, 2006).

Complex network theory considers air travel systems as networks, where airports are nodes and flight routes are edges. Network analysis and robustness studies have been performed by researchers in order to understand the network's dynamic characteristics. Indicative examples in the literature include the work of Guimerà and Amaral (2004), Guimerà et al. (2005), and Xu and Harriss (2008). Due to the dynamic nature of such systems, temporal analysis can provide insight on factors affecting their operations (Gillen and Morrison, 2005). Temporal evolution of airport networks is assessed by treating the evolving network as a series of static snapshots, each one representing a complex network for that specific time instance (da Rocha, 2009; Zhang et al., 2010; Lin and Ban, 2014; Jia et al., 2014; Sun et al., 2015; Wandelt and Sun, 2015; Wandelt et al., 2017; Dai et al., 2018). An extensive review of the research is provided by Rocha (2017).

The aforementioned studies suggest that air transportation networks are usually small-world networks with power-law decaying degree and betweenness distributions, corresponding to scale-free networks. Moreover, the most connected cities, (cities with the highest degree), were found to differ from the most central cities (cities with the highest betweenness), due to economic and geo-political constraints (Guimerà and Amaral, 2004; Lin and Ban, 2014). The U.S. ATN in particular displays strong disassortative mixing patterns (thus a well-developed hub and spoke (HS) structure), being among the most efficient and mature networks in the world (Xu and Harriss, 2008; Wandelt et al., 2017). Temporal studies of the U.S. ATN indicate that the restructuring which occurred after 2001 had a huge impact on its structural features (Jia et al., 2014; Lin and Ban, 2014).

Despite the vulnerability of such networks to intentional attacks against critical hubs (Barabási and Albert, 2002; Crucitti et al., 2003), very few researchers have addressed the relationship between ATN topology and its tolerance to significant disruptive events. Organizational complexities should be accounted for during a robustness evaluation, thus rendering a complex network analysis necessary (Lordan et al., 2014a). Chi and Cai (2004) used complex network theory to study the effect of random failures and intentional attacks on the topological properties of the U.S. ATN. Intentional attacks were simulated by a node deletion method, similar to the edge deletion method of Girvan and Newman (2002). Wuellner et al. (2010) and Lordan et al. (2014b) proved that the effects of targeted attack based on betweenness centrality, rather than by degree can be more destructive for the network. In 2015, Janić suggested a resilience indicator to express the network's operational level under given conditions. Zhou et al. (2018) proposed a weighted efficiency to assess the ability of an ATN to maintain its performance when facing random airport closures. It should be noted that all said investigations have not accounted for the network's temporal evolution. A literature survey of studies focused on ATN robustness can be found in (Lordan et al., 2014a).

Evidently, the topological properties of air transportation networks worldwide have been extensively examined from a complex network perspective. On the other hand, the robustness of such networks still remains largely unexplored, with few publications focusing on the system's vulnerability to targeted attacks of the most central nodes. Furthermore, to the authors' knowledge, a temporal study of the robustness of air transportation networks and their tolerance to targeted attacks, using network analytics theory has not been performed in the literature. The temporal evolution of ATN topological properties can identify weaknesses inherent to the network structure or emerging naturally as a result of the network's organic growth, thus allowing to define future operational strategies and policies.

# 3    Data and methodology

In the following section, the input dataset is outlined. Subsequently, the basic network theory is described, and all topological parameters used in this study are concisely defined. The network analysis is performed based on data collected by the U.S. Bureau of Transportation statistics (BTS), and is executed in MATLAB. The developed code for the simulations has been made publicly available (on https://github.com/vub-dl/atn-na) to facilitate reproduction of the results and application to alternative networks.

## 3.1 Dataset

The study of the U.S. ATN is realized based on the dataset provided by the U.S. BTS, which can be obtained through the official website (https://www.transtats.bts.gov/). Total annual commercial domestic flight data are downloaded for each year between 1996 and 2016. The input raw data comprises of airport origin and destination information, number of flights performed between the given airport pair, and number of passengers flown.

The U.S. ATN is modelled at each time as a binary directed network. A complex network analysis methodology is applied, in which nodes represent airports and edges represent flight route connections. Two airports are considered to be connected if at least one non-stop commercial airline route exists between them.

## 3.2 Topological properties

Several parameters are used to identify the most connected and central nodes of the network, in order to conduct further research on its global topological properties. Subsequently, network resilience and tolerance to intentional attack are evaluated.

### 3.2.1 Centrality measures

The centrality of nodes, thus the identification of nodes which are more central than others is of critical importance in network analytics. In the present study, node centrality is evaluated based on node degree and betweenness. According to Freeman (1978), the degree of a node is equal to the number of adjacencies in the network, thus the number of nodes that node is connected to. This measure can be generalized as

$$k_i = \sum_{j=1}^{N} e_{ij} \quad (1)$$

where $i$ is the focal node and $j$ represents all other nodes in the network. $N$ is the total number of nodes in the network and $e_{ij}$ defines the elements of the adjacency matrix. The topology of the network can be represented by an adjacency matrix, where the elements of $e_{ij}$ are 1 if $i$ and $j$ are connected, and 0 otherwise. A large degree indicates a well-connected node.

According to Barrat et al. (2004), node strength is defined as

$$s_i = \sum_{j=1}^{N} e_{ij} w_{ij} \quad (2)$$

where $w_{ij}$ is the weight of the connection between nodes $i$ and $j$. The degree defines the number of flight route connections of an airport, whereas strength indicates cumulative traffic handled through that airport.

Weights can be based on total flights $w_{ij}^F$ or total passengers $w_{ij}^P$. Evidently, the airport degree measures how connected the airport is, whereas its strength indicates how travel intensive it is. In the current study global statistics are sought, since they are of importance from a national security and an air traffic management perspective. Hence, an unweighted analysis is considered sufficient.

Airports as nodes are almost always connected in both directions, hence the indegree and outdegree (degree based on inbound and outbound flights, respectively) of a given node are almost equal. Therefore, in the present study all edges are considered as undirected. The sum of indegree and outdegree is considered as the total degree.

For an aviation network, betweenness centrality is important to quantify the bridge importance of an airport to the system (Guimerà et al., 2005). Betweenness is indicative of the probability that a node is in the shortest paths between all node pairs in the network (Freeman et al., 1979). In a binary network, the shortest path $d_{ij}$ between nodes *i* and *j*, is defined as the shortest geodesic distance between any two nodes. It should be noted that $d_{ij}$ is an indicator of geodesic distance, whereas the spatial distance will be denoted as $D_{ij}$. The geodesic distance $d_{ij}$ is adopted for shortest path calculation, since in the airline industry, passengers care more about the number of connections they need from origin to destination (Latora and Marchiori, 2001). A necessary assumption is that intermediary nodes lead to increased interaction costs and thus the shortest path involves the least number of intermediary nodes. According to Freeman et al. (1979), betweenness centrality is expressed as

$$B_i = \frac{1}{(N-1)(N-2)} \sum_{i \neq j \neq k} \frac{g_{jk}(i)}{g_{jk}} \qquad (3)$$

where $g_{jk}$ is the number of binary shortest paths between two nodes, whereas $g_{jk}(i)$ is the number of those paths that pass through node *i*.

### 3.2.2 Network assortativity

Network assortativity can be used to assess whether a given network follows a point to point (PP) or HS paradigm. The tendency of high-degree nodes to connect to other high-degree nodes is proof of PP (hence assortative) structures, whereas high-degree nodes tending to connect to low-degree nodes is an indication of HS (disassortative) structures.

In PP connectivity, passengers take direct non-stop flights, usually covering a short distance. Point to point networks reduce travel time, thus being valued by travelers. They require however large markets for connection of large city pairs, in order to be economically viable (Cook and Goodwin, 2008). Hub and spoke systems typically enable passengers to travel non-stop only to a few hubs, from which they can subsequently transit to their final destination. They minimize the number of connections, instead of overall distance travelled (Gastner and Newman, 2006) thus requiring the fewest aircraft and allowing airlines to exploit economies of scale and scope. Despite the obvious advantages, main shortcomings of HS systems lie in their complexity and operational cost. Moreover, they are highly susceptible to delays, since delays can propagate over the network. In a HS network, hubs are the most critical and undoubtedly vulnerable nodes of the system (Cook and Goodwin, 2008).

In order to quantify the network's assortativity, the Gini coefficient is chosen. It has been commonly used in economics as a measure of inequality (Sen, 1973) and has been further applied for characterization of airline traffic patterns (Reynolds-Feighan, 1998; Wuellner et al., 2010). It has been preferred over the assortativity coefficient for measuring assortative structures, since the latter is very sensitive to outliers (Devlin et al., 1975) which can introduce erroneous artifacts (Wuellner et al., 2010). Such outliers are present in airline networks (i.e. superhubs).

The Gini coefficient, $G$, is defined as

$$G = \frac{\sum_{i=1}^{N}\sum_{j=1}^{N}|k_i - k_j|}{2MN} \tag{4}$$

where $M$ is the total number of edges in the network. It allows measuring the magnitude of node degree difference between each node pair in the network, normalized by average node degree.

### 3.2.3 Network efficiency

Network efficiency $E$ can be defined as a global parameter by assuming that communication efficiency between nodes depends on the shortest path length. According to (Latora and Marchiori, 2001):

$$E = \frac{1}{N(N-1)} \sum_{i \neq j} \frac{1}{d_{ij}} \tag{5}$$

This is an indicator or local and global efficiency and measures how efficiently information is exchanged within a network.

### 3.2.4 Small-world properties

The network is examined aiming to identify its small-world properties. A small-world property is indicative of a network whose nodes can be reached by a small number of connections (Watts and Strogatz, 1998). The existence of cliques, hence sub-networks which have connections between almost any two nodes within them is such an indicator. To this end, the clustering coefficient and the characteristic path length are introduced. Mathematically, small-world properties are expressed by a large clustering coefficient and a small characteristic path length.

The clustering coefficient $C_i$ represents the possibility that two neighbors of a node may be themselves connected, thus quantifying the inherent tendency to cluster:

$$C_i = \frac{2m_i}{k_i(k_i - 1)} \tag{6}$$

It measures how interconnected a neighborhood is, based on the interconnected triples in the said neighborhood. The clustering coefficient ranges between 0 (if no ties exist between the neighbors) and 1 (if all possible ties exists). Here, $m_i$ is the number of edges connecting the neighbors of node $i$ (Watts and Strogatz, 1998). The global clustering coefficient of a network is equal to the mean of the clustering coefficients of all nodes in the network.

The characteristic path length is additionally introduced to characterize the average shortest-path length between any two airports in the system:

$$L = \frac{1}{N(N-1)} \sum_{i \neq j} d_{ij} \tag{7}$$

It quantifies the average of least possible connections between any two airports in the network. Evidently, a network of a small characteristic path length is associated with high efficiency (Eq. (5)), where passengers can conveniently reach their destination through few connections.

### 3.2.5 Scale-free properties

Topology study of the ATN has shown that it is a scale-free network, following a truncated power-law distribution (Guimerà and Amaral, 2004; Guimerà et al., 2005). Scale-free networks are free of a characteristic scale. Therefore, the degree distribution $P(k)$ decays as a power-law, that is $P(k) \sim k^{-\gamma}$, where $\gamma$ is the exponent (Barabási and Albert, 2002).

## 3.3 Network robustness

Network robustness is defined in this work as a combination of the network's resilience to randomly occurring disruptive events and its tolerance to a targeted attack.

### 3.3.1 Network resilience

The resilience of an ATN can be defined as the network's ability to maintain its operations at the required safety levels, during a disruptive event. Resilience does not relate to the aftermath of a given event, but rather to the immediate impact (Chen and Miller-Hooks, 2012). Disruptive events can be identified as bad weather conditions, natural disasters, network infrastructure failures, personnel strikes, accidents, and terrorist attacks. Based on the magnitude of such an event, closure of one or more airports may lead to flight delays and/or cancellations, which may in turn propagate and affect other nodes.

The network's resilience to a large-scale disruptive event is assessed based on the methodology suggested by Janić (2015) and is indicative of the network's ability to sustain its operability during the impact of a given disruptive event. The author empirically validated their approach for air transportation systems by applying it to demonstrate how hurricane Sandy compromised the resilience of the network on the north-east U.S., during October 2012.

According to the model of Janić (2015), the relative importance of each airport should be considered first. Relative strength based on the total number of inbound and outbound flights $\hat{s}_i^F$ (obtained from Eq. (2)) is computed at each node:

$$\hat{s}_i^F = \frac{s_i^F}{\sum_{j=1}^{N} s_j^F} \tag{8}$$

The relative strength is indicative of airports operating at regular conditions and at the nominal flight capacity ratio.

Subsequently, the self-excluding importance of a given airport is then defined by assuming that its other connected airports do not include it:

$$\upsilon_i = \frac{s_i^F}{\sum_{j=1}^{N} s_j^F - 2 s_i^F s} \tag{9}$$

Finally, resilience of a given airport can be calculated as

$$R_i = \sum_{j=1/ j \neq i}^{N} \upsilon_j \left( \delta_{ji} w_{ji}^F \right) \tag{10}$$

with $\delta_{ji}$ representing the Kronecker delta. Evidently, resilience is proportional to the sum of the product of the self-excluding importance and the number of total flights realized to and from each connected airport. The global resilience of the network is then defined as the sum of the weighted resilience of each airport (node) belonging to the said network:

$$R = \sum_{i=1}^{N} \hat{s}_i^F R_i \tag{11}$$

### 3.3.2 Targeted attack tolerance

A node deletion method (Chi and Cai, 2004) is applied to assess the network's behavior to airport closure after a targeted attack. This approach is based on the edge deletion algorithm of Girvan and Newman (2002), which was first used for community finding in social and biological networks. According to Girvan and Newman (2002), the edges with the highest betweenness connect communities. As such, removal of those edges allows groups to be separated from one another, in order to reveal the network's underlying community structure. Targeted removal is preferred to provide insight on the worst-case scenario. Although the probability of such an incidence is low, the magnitude of the impact on the system would be maximized (Lordan et al., 2014b). Hence, the most critical nodes are targeted by simulated intentional attacks.

Betweenness centrality is used as the criterion of node importance, since targeting by betweenness has shown to cause a more rapid breakdown of the network, compared to other measures (Wuellner et al., 2010; Lordan et al., 2014b). Betweenness allows taking advantage of the multi-community structure of airport networks, by selecting strategic hubs. Removal of these hubs can result in the disconnection of whole geographical regions (Lordan et al., 2014b).

The adaptive procedure applied here starts with the detection of the node with the highest betweenness, which is removed. Subsequently, the centrality measures are recomputed for the updated network and the node with the highest betweenness is removed once again. This procedure is realized for 50 iterations. The topological measures used to assess the impact of node deletion is the size of the largest connected component $S$, as well as the network efficiency $E$.

# 4 Results and Discussion

In this section, the results obtained from the network analysis are presented and their temporal evolution is discussed. Subsequently, the small-world and scale-free properties of the network are investigated, along with the presence of assortative or disassortative structures. Finally, the robustness of the system is evaluated based on the evolution of resilience, efficiency, and vulnerability to targeted attacks.

## 4.2 Network analysis

Initially, results obtained using network analytics on the U.S. ATN between 1996 and 2016 are presented. Total passenger and flight data is acquired yearly. Thus, each year is considered as a snapshot, finally leading to 21 snapshots. Each snapshot is then analyzed as a separate network.

Centrality measures are initially computed and examined on a global basis. Small-world and scale-free characteristics are additionally identified and discussed. Finally, the evolution of assortative structures in the network is addressed.

### 4.2.1 Centrality measures

The network's basic centrality measures for each year, are illustrated in Figures 2 and 3. Figure 2 displays the total number of network nodes (airports) and edges (routes), whereas Figure 3 depicts the average flight distance and average node betweenness plotted against time.

The evolution of overall nodes and edges in the network follows similar trend (Figure 2). The most outstanding point in time is year 2002, right after the events of 9/11, when major restructuring occurred. In 2002, 538 new airports and 8150 new routes were added to the network, aiming to improve its robustness (Lin and Ban, 2014, Jia et al., 2014). After 2002, some growth is noted, finally reaching a global maximum in 2007, for both nodes and edges. Subsequently, the 2008 recession is evident. Dobruskzes and Van Hamme (2011) mention that during January 2008 and 2010, 739 routes were suppressed and only 349 new routes were created. Since 2013, network expansion recovers, eventually resulting in a total of 1217 airports and 26060 routes, in 2016.

Growth and expansion of the U.S. ATN is corroborated by the increase observed in nodes and edges, over time.

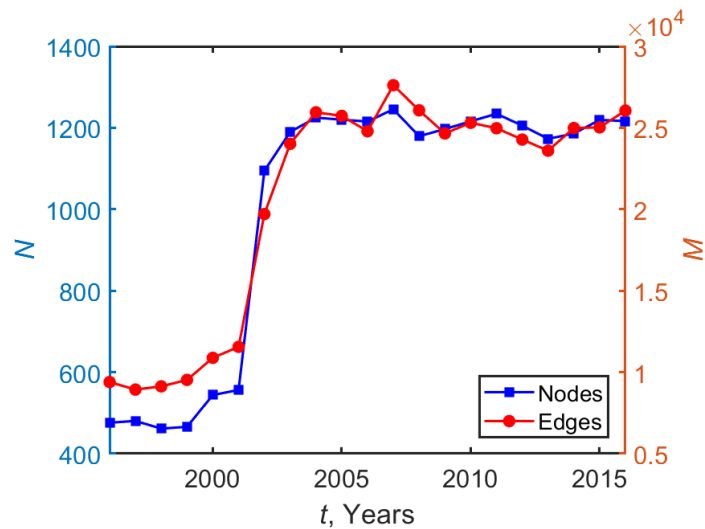

*Figure 2.* Temporal evolution of the total nodes (*N*) and edges (*M*) in the U.S. ATN, between 1996 and 2016.

Figure 3 depicts the temporal variation of the average flight distance $\langle D \rangle$, plotted alongside the calculated average node betweenness $\langle B \rangle$. Average quantities refer to the global mean among all nodes in the network. Average flight distance displays a generally increasing tendency which was interrupted by the network's restructuring in 2002 (Lin and Ban, 2014). Slow growth is observed thereafter, with a local minimum at 2009 amidst the recession. Since 2010, the average flight distance increases reaching a global maximum of 1124 km in 2016.

On the other hand, the curve of average node betweenness initially shows declining trend. A drastic increase is observed in 2002 when more airports were introduced and airlines added new short-haul routes (see Figure 2). The increase by almost 30% in average betweenness from 2001 to 2002, coincides with the sharp decline of flight distance (Figure 2). The average betweenness reaches a global maximum in 2002 and since then decreases, with a peak amidst the recession. Its value eventually drops below 2 in 2016. This can be considered an indication of growing robustness, since an even distribution of betweenness centrality among nodes indicates higher tolerance to intentional attacks (Lin and Ban, 2014).

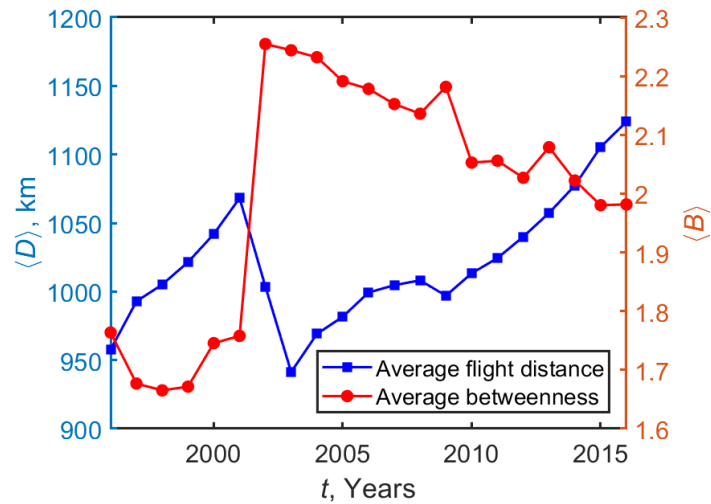

*Figure 3.* Temporal evolution of average flight distance and average node betweenness in the U.S. ATN, between 1996 and 2016.

Further insight into the average flight distance is provided in Figure 4, where the absolute values of short-haul (lower than 700 miles) and long-haul (greater than 700 miles) flights are plotted. The ratio of short-haul over long-haul flights is also displayed. This graph indeed verifies the introduction of several short distance flights in 2002, resulting in almost doubling of the short-haul over long-haul flight. After the initial restructuring, the ratio tends to converge to the values observed before 9/11.

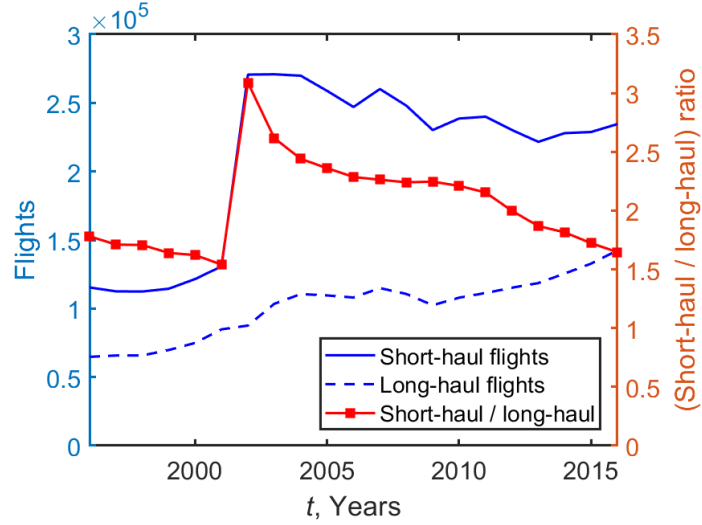

*Figure 4.* Temporal evolution of short- and long- haul flights in the U.S. ATN, between 1996 and 2016. Flights over distance lower than 700 miles are considered as short-haul, whereas flights over distance greater than 700 miles are considered as long-haul.

One may conclude that the increase of average flight distance between 2002 and 2016, is directly related to the decreasing average betweenness during the same period. With the addition of intermediate airports, airlines can exploit HS structures to introduce non-direct routes.

### 4.2.2 Small-world properties

The evolution of the network's small-world properties is examined from a global perspective. The variation of the global average clustering coefficient and global average shortest path length is illustrated in Figure 5.

The average clustering coefficient displays a global maximum in 1996 and eventually reaches its global minimum in 2001. On the other hand, the average shortest path length reaches its global minimum during 1999 and its global maximum in 2002. Hence, the 2001 events have a critical impact on small-world properties. Subsequently, both parameters show a generally declining trend. In 2016, the values of $\langle C \rangle$ and $\langle L \rangle$ are 0.42 and 2.85, respectively, being indicative of a small-world network. On average, a node can be reached by any other node in the network by less than 3 intermediate hops.

Evidently, small world properties are found in the domestic U.S. ATN. The U.S. ATN shows a decline of average path length, hence evolving into a more efficient network over time. The increase of average flight distance (see Figure 3) implies that passengers can fly further by fewer hops. Moreover, the U.S. ATN displays a tendency to be gradually less clustered. This can be associated with the generally declining betweenness seen in Figure 3, which leads to the central weight being shared among nodes.

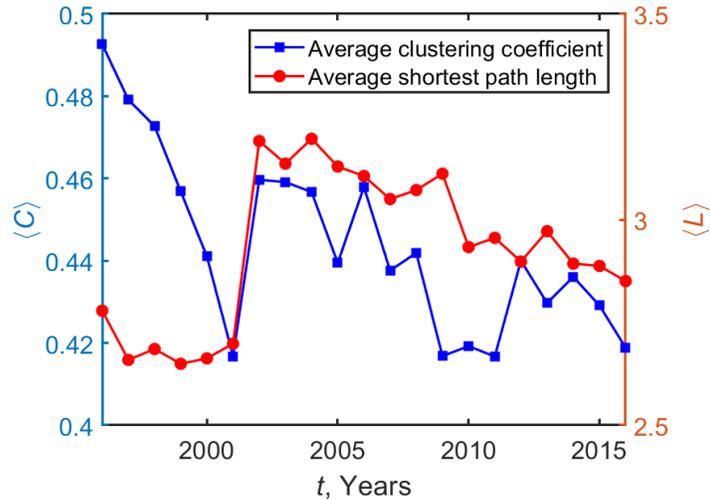

*Figure 5.* Temporal evolution of small-world properties in the U.S. ATN, between 1996 and 2016.

### 4.2.3 Scale-free properties

Main characteristic of scale-free networks is the existence of hub nodes with high connectivity, whereas most nodes depict rather few connections. Such behavior was intentionally forced to the U.S. ATN during the restructuring of 2002.

Cumulative distributions of degree and betweenness are plotted and analyzed, in order to identify scale-free properties. We analyze the distribution of in-degree ($k_{in}$) and out-degree ($k_{out}$), with $p(k)$ the observed probability of the degree of a given node to have value *k*. Since raw probability distributions are noisy, the cumulative distributions are constructed according to $P(k) = \sum_{i>k} p(k)$.

Figure 6 shows the plots of cumulative in-degree and out-degree distributions with a time interval of 5 years. The curves follow a power-law distribution with truncations. A similar trend is observed between every time snapshot. These conclusions agree with the results of Xu and Harriss (2008), Lin and Ban (2014), and Jia et al. (2014). The power-law behavior of passenger distribution suggests high heterogeneity of air traffic (Xu and Harriss, 2008).

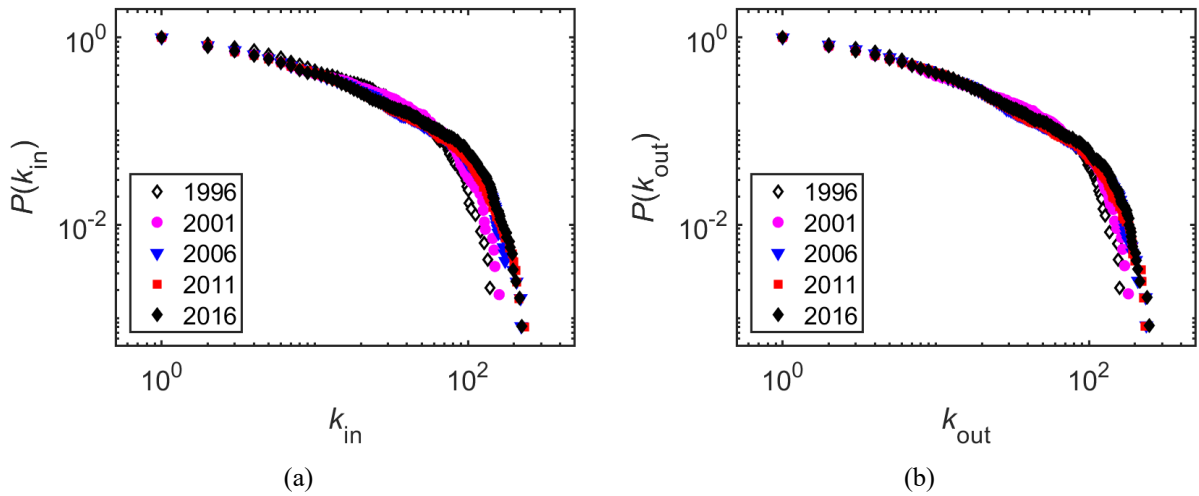

Figure 6. The evolution of the cumulative in-degree $P(k_{in})$ and out-degree $P(k_{out})$ distribution on log-log scales.

The cumulative betweenness distribution is presented in Figure 7, also decaying as a power law, with a large exponent value. This is a common characteristic among air transportation networks and implies anomalously large betweenness centralities (Guimerà et al., 2005; Wuellner et al., 2010), as specific airports which cannot be considered hubs display a very large betweennesses (Guimerà et al., 2005).

Scale-free characteristics of the U.S. ATN show some convergence during recent years. Hence the network retains its scale-free properties.

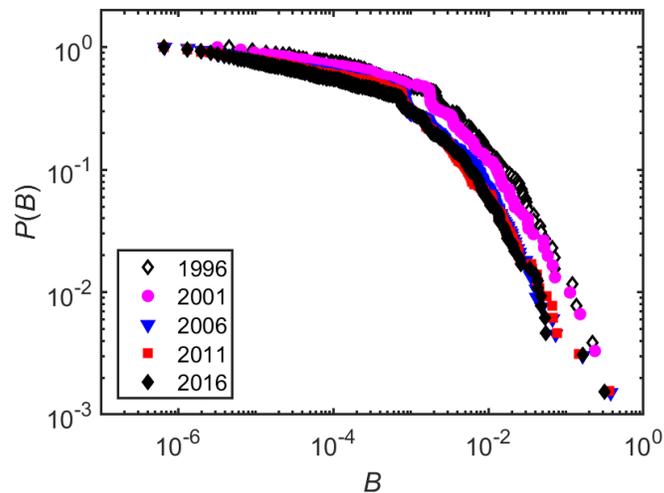

Figure 7. The evolution of the cumulative betweenness $P(B)$ distribution on log-log scales.

### 4.2.4 Assortativity

The temporal evolution of the network's assortativity is evaluated through the Gini coefficient, which generally displays an increasing trend, hence encouraging the development of HS structures. The minimum value of 0.64 appears in 1996, while the global maximum value is just below 0.7 and occurs at the most recent time. The U.S. ATN is thus highly disassortative, with increasing disassortative features over time. The two local minima seen in Figure 8 coincide with the 9/11 crisis and the 2008 recession. The former can be explained by the restructuring, as well as the dehubbing of specific airports in the immediate aftermath of the attacks (Redondi et al., 2012). The contribution of HS structures to the network's restructuring is corroborated by the findings of Figure 3. The aviation fuel price surge between 2004 and 2008 amplified the growth of HS structures, since connections to non-hub airports serving small communities were most sensitive to fuel price increases. Non-hub airports lost approximately 12% of their connections during this period (Morrison et al., 2010).

Overall, there is an evident trend towards stronger HS structures in the U.S. ATN over time (Jia et al., 2014).

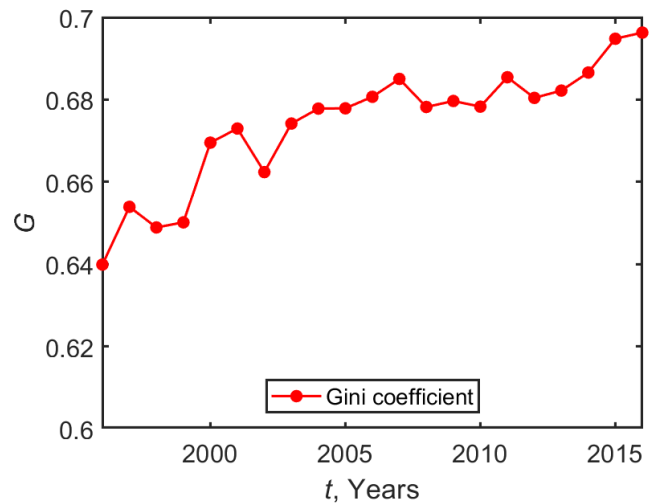

*Figure 8.* Temporal evolution of the Gini coefficient of the U.S. ATN, between 1996 and 2016.

### 4.3 Network robustness

Based on the initial network analysis a temporal robustness study is performed. The temporal variation of the efficiency and resilience of the U.S. ATN are initially evaluated. Subsequently, the network's targeted attack tolerance is estimated. Due to the scale-free properties observed in section 4.2.3, targeted removal of the most critical nodes is realized.

### 4.3.1 Network efficiency and resilience

Figure 9 presents comparative plots of network efficiency and resilience. Both curves display their global maxima a few years before the events of 2001 and show a steep drop between 2001 and 2002. During that period, efficiency and resilience experience a drop of almost 20%. Network resilience reaches its global minimum in 2003. The curve only shows notable growth again after the wake of the 2008 economic recession. The system's resilience has presented a continuously increasing trend since. The network's efficiency has shown a continuous decrease of efficiency since 2008. This small decrease of efficiency does not necessarily imply a degradation of the system, due to the continuous increase of nodes (airports) and edges (routes) throughout the network during recent years (Lin & Ban, 2014). Efficiency is evidently inversely related to the average shortest path length (see Figure).

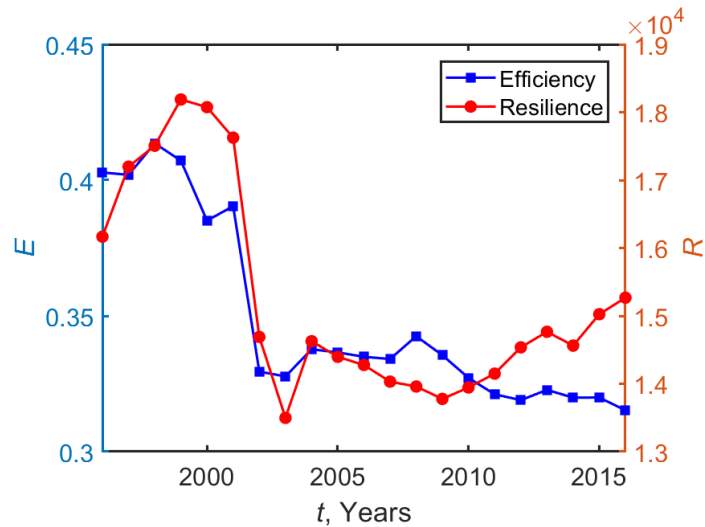

*Figure 9.* Temporal evolution of the efficiency and resilience of the U.S. ATN, between 1996 and 2016.

### 4.3.2 Targeted attack tolerance

In order to assess the network's tolerance to a targeted attack, we start by detecting the most central node, hence the node with the highest betweenness. Since the node isolation approach is adaptive, the said node is removed and the centrality measures are recomputed for the updated network. This procedure is repeated for 50 iterations.

In Figure 10, the impact of targeted node removal is displayed with a time interval of 5 years. The monitoring variables are the network's largest connected component $S$ (Figures 10 (a) and (c)) and the efficiency $E$ (Figures 10 (b) and (d)).

Figures 10 (a) and (b) illustrate the findings after the iterative removal of 50 airports, which leads to the network retaining 86% of its structure (Figure 10 (a)). This is certainly an improvement, compared to the network in 1996 and 2001. In 2006, the U.S. ATN displays the best tolerance among the five snapshots, whereas between 2011 and 2016, convergence is observed. Removal of 50 nodes results in reduced network efficiency by more than 40% for all time instances (Figure 10 (b)). Figures 10 (c) and (d) are zoomed in on the 10 first iterations, thus clearly depicting the immediate impact of a targeted attack. Removal of the most connected node has the most severe effects in year 2016 (see Figures 10 (c) and (d)). Removal of the second most connected node has the most critical impact for year 2016, in regard to $S$ (Figure 10 (c)). Generally, the snapshot corresponding to 2016 displays the most rapid breakdown of $S$ and $E$, being only less vulnerable than 2001. This is proof of the system's vulnerability to targeted attacks becoming more prominent over time.

Further insight on the restructuring following 9/11 is provided in Figure 11, for the period between 2000 and 2004. Figures 11 (a) and (b) display $S$ and $E$ results after the iterative removal of 50 airports, generally illustrating the abrupt change of the network's characteristics after the restructuring. Notable convergence is observed between 2002 and 2004, and thus an improved tolerance compared to the network's state before 9/11. Figures 11 (c) and (d) are zoomed in on the 10 first iterations, proving that the most vulnerable structure is generally observed during year 2000. The restructuring resulted in overall improvement of the system's tolerance to targeted attacks, since the curves corresponding to 2004 display the most robust characteristics.

Evidently, removal of small fraction of selected nodes can cause severe damage to the functioning and operations of the U.S. ATN, as expected by its scale-free properties (Chi and Cai (2004), Wuellner et al. (2010), and Lordan et al. (2014b)). Due to the multi-community structure of the U.S. ATN, removal of specific hubs can result in the disconnection of whole geographical regions (e.g. the Alaska region, which connects to the rest of the U.S. mainly through Anchorage and Fairbanks airports) (Lordan et al., 2014b). The discussed results suggest that the system's robustness does not show improvement over time. Moreover, the system in 2016 shows to be the most vulnerable to a targeted attack on removal of the first few most connected nodes since the 2001 restructuring (Figures 10 (c) and (d)).

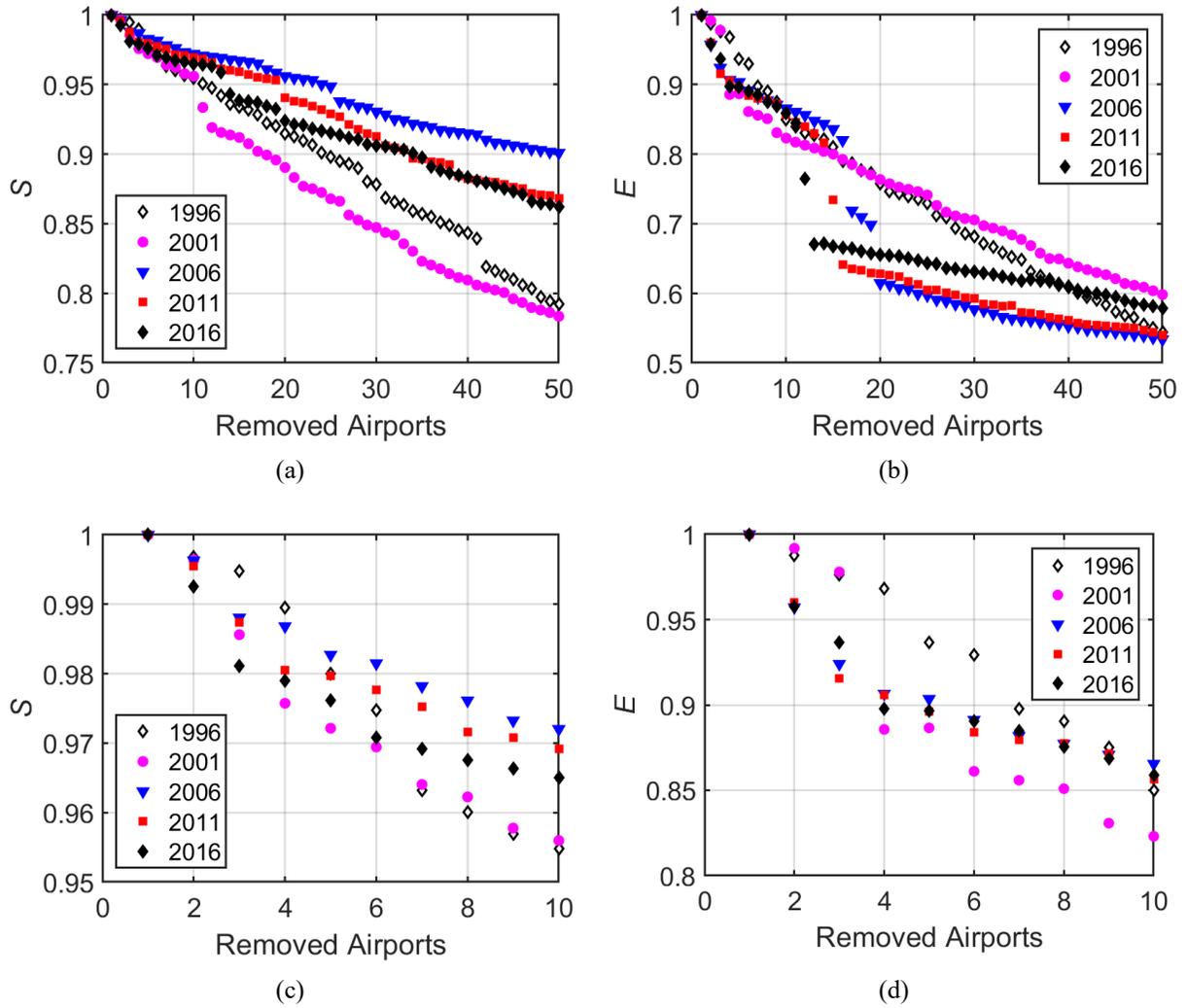

*Figure 10.* The evolution of the network's largest connected component *S* and efficiency *E* as a function of nodes removed. Parameters *S* and *E* are normalized by their respective maximum value observed for each specific year.

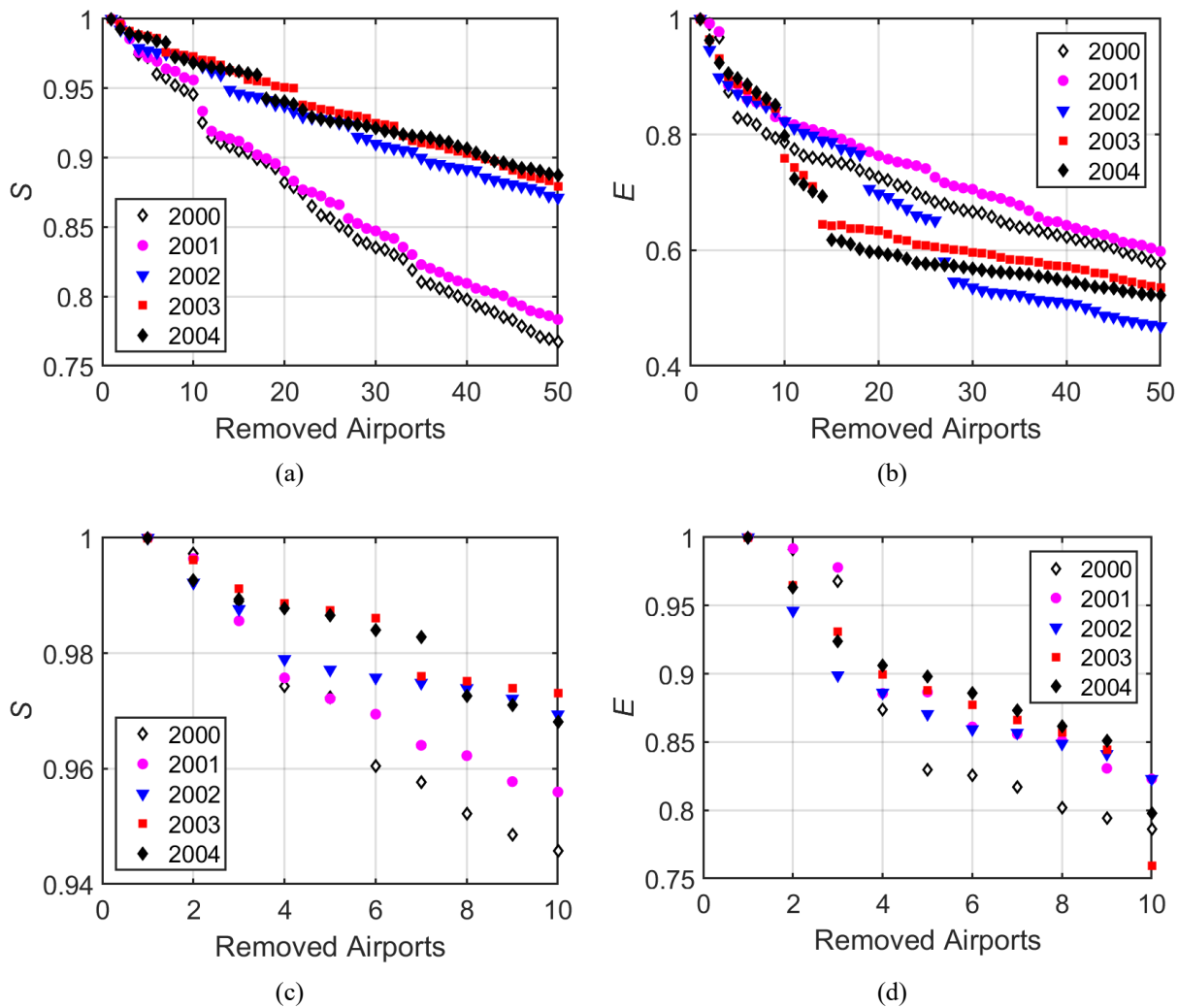

*Figure 11.* The evolution of the network's largest connected component *S* and efficiency *E* as a function of nodes removed, during the restructuring period. Parameters *S* and *E* are normalized by their respective maximum value observed for each specific year.

## 4.4 Discussion

Based on the results of the performed network analysis, clear observations can be initially made on the state of the U.S. ATN immediately after the 9/11 attacks, as well as during the short-term restructuring that took place. The long-term evolution of the system's topological properties and robustness provides important information for the current state of the U.S. ATN and allows defining guidelines for future action.

The direct impact of the attacks in 2001 was rather limited (see Figures 5, 8, and 9), since data from this year only captured the system's reaction to increase security and recover operations. Some routes were eliminated and this resulted in suboptimal operation of the network and lower traffic overall.

The most severe shocks are observed in 2002. The restructuring resulted in the introduction of new airports, which were connected to the already well-connected airports of the system, aiming to construct a more connected network (Lin and Ban, 2014). This is reflected in the drastic increase of average node betweenness and the steep decline of average flight distance in 2002 (Figure 3). A sharp expansion of flights, nodes, and edges, during that same year, also points to this direction (Figure 2). Airlines had already replaced several large aircraft with smaller ones. This rendered them more flexible and allowed them to address the eventual growth of passenger demand, reaching in the meantime smaller markets. Major airlines could thus maintain the scope and scale of their operations, while preserving their market share against low cost carriers. Passenger traffic began its recovery with some delay, mainly due to passengers being wary of air travel and discouraged by security inspections. The severity of the 9/11 events is captured by the steep decline of efficiency and resilience in Figure 9. Although the efficiency decline should not be considered as a degradation of the system when growth of nodes and edges is significant (Lin and Ban, 2014), the drop of resilience is an alarming indicator. The restructuring additionally improved the system's tolerance to targeted attacks (Figure 11).

Since 2002, much smoother and generally monotonic trends are observed. However, the general volatile state of the market and particularly the 2008 recession had a toll on the network's expansion. Despite passenger and flight traffic being affected, the impact on all other network statistics was of a much lower magnitude than the 2001 events.

The most recent observations, provide a view of the current state of the U.S. ATN. The network displays continuous growth in passenger and flight traffic, airports, and routes (Figures 1 and 2). Average flight distance increases, whereas average betweenness decreases (Figure 3). The increase of available intermediate airports, enables airlines to exploit HS structures in order to introduce non-direct routes and increase route efficiency (Figure 8). Furthermore, the network is retaining its small-world properties, despite reduction of clustering. The average path length also displays a declining trend, (as expected by the increasing average flight distance) which is indicative of an efficient system (Figure 5). On the other hand, the robustness of the U.S. ATN does not show improvement, since it maintains its scale-free properties (Figures 6 and 7) and thus its vulnerability to targeted attacks. An alarming characteristic is that its tolerance to a targeted attack scenario has deteriorated, based on the most recent data (Figure 9). Some positive conclusions on the network's robustness comprise of its increasing resilience over time (Figure 9) and the decreasing betweenness (Figure 3), which are desirable in robust networks.

Overall, the 9/11 terrorist attacks triggered an unprecedented restructuring of the U.S. ATN. Airlines reconsidered their strategy and optimized their operations to survive, thus allowing the network to recover rapidly and become even more efficient. The devastating effects of the 9/11 events called for the establishment of severe security measures. As such, in the short-term period after 2001, the network's tolerance showed undoubtable improvement. Since then, the U.S. ATN has continued its evolution in a dynamic manner, being generally volatile and susceptible to global economic fluctuations. One may conclude that in recent years, the network's global statistics and topological properties display continuous growth, as the network continues to expand. The robustness of the system, on the other hand, has not shown any significantly improving trend, despite the increase of its resilience. The system's tolerance to targeted attacks has displayed deteriorating tendencies during recent years. A concise outline of the examined network parameters is provided in Table 1.

| Parameter | Range | Min. | Mean | Max. | Description |
|---|---|---|---|---|---|
| Nodes, $N$ | ~ + | 461.5 | 1,000.1 | 1,245.5 | Number of airports |
| Edges, $M$ | ~ + | 8,934 | 20,582 | 27,617 | Flight route connections |
| Avg. flight distance, $\langle D \rangle$ | ~ + | 941.3 | 1,020.7 | 1,123.9 | Average flight distance in km |
| Avg. node betweenness, $\langle B \rangle$ | ~ + | 1.665 | 2.002 | 2.255 | Increase indicates higher bridge importance of nodes, thus vulnerability to intentional attacks |
| Avg. clustering coeff., $\langle C \rangle$ | [0,1] | 0.417 | 0.444 | 0.493 | Increase indicates increasing small-world properties, thus higher network efficiency |
| Avg. shortest path length, $\langle L \rangle$ | ~ + | 2.650 | 2.929 | 3.196 | Decrease indicates increasing small-world properties, thus higher network efficiency |
| Gini coeff., $G$ | [0,1] | 0.640 | 0.674 | 0.696 | Increase indicates stronger HS structures |
| Efficiency, $E$ | [0,1] | 0.315 | 0.349 | 0.413 | Inversely related to average shortest path length |
| Resilience, $R$ | ~ + | 1.35e+04 | 1.53e+04 | 1.82e+04 | Indicates network ability to sustain operational levels |

*Table 1.* Summary of the computed variables.

# 5 Conclusion

Due to the U.S. air transportation network's criticality to the mobility and functioning of local economies, it is imperative to assess its topological evolution and ensure that it is well-connected, efficient, and robust. In the present paper, a study of the domestic U.S. ATN was realized between 1996 and 2016, to evaluate the temporal evolution of its topological properties and robustness. Network analytics were used to evaluate the system's topological properties. Emphasis was placed on the effects of restructuring that followed the 9/11 events. Centrality measures were computed. Global resilience of the system was subsequently evaluated. A node deletion method (Chi and Cai, 2004) was also applied to assess the network's tolerance and its capacity after removal of critical nodes, thus simulating the worst-case scenario of a targeted attack. The network's efficiency and largest connected component were monitored to assess the impact of such an event.

The main findings of this study showed that the 9/11 terrorist attacks triggered vast restructuring of the network. Commercial airlines improved their efficiency and the U.S. government focused on security. The network thus recovered rapidly and became even more efficient. The devastating effects of the attacks themselves, exposed security weaknesses, resulting in the immediate establishment of severe security measures. Therefore, in the short-term period after 2001, the network's tolerance showed significant improvement. Since then, the U.S. ATN has continued evolving, displaying continuous growth and expansion. However, the robustness of the system, has generally not shown any significantly improving tendency. Although the system's ability to sustain its operational level under extreme circumstances has improved, its tolerance to targeted attacks has recently displayed signs of deterioration.

The deteriorating characteristics of U.S. ATN robustness highlight the need for future improvement and restructuring. Since demand for air travel is projected to grow in years to come, a well-connected, efficient, and robust network is imperative to maintain growth. In order to ensure sustainable growth of the U.S. ATN a complex network analysis combined with an economic analysis is suggested to identify the most critical nodes. The vulnerability of these nodes should be evaluated, along with the impact of their removal. Future restructuring should take these findings into account in order to protect the critical nodes and establish fast response mechanisms. Furthermore, minimizing the scale-free properties of the ATN would render it less vulnerable to targeted attacks. Other suggestions include the removal of few of the smallest airports and encouraging point to point structures. Moreover, future restructuring should consider the necessity for expansion of the current airport network to accommodate the increasing passenger demand, as well as all constraints that limit this expansion.

We have demonstrated the applicability of the overall methodology on the U.S. ATN. The methodology can in turn be extended to different transportation networks, in order to provide a general perspective of a system's vulnerabilities. Weights can be also easily accounted for, depending on the different regions of transportation modes. The developed code has been made publicly available to facilitate reproduction of the results and application on alternative networks.

# Bibliography

Anderson, R., Fraser, C., Ghani, A., Donnelly, C., Riley, S., Ferguson, N., Hedley, A. (2004). Epidemiology, transmission dynamics and control of SARS: the 2002–2003 epidemic. *Philosophical Transactions of the Royal Society of London. Series B: Biological Sciences, 359*(1447), 1091-1105.

Barabási, A., & Albert, R. (2002). Statistical mechanics of complex networks. *Reviews of Modern Physics, 74*(1), 47-97.

Barrat, A., Barthélemy, M., Pastor-Satorras, R., & Vespignani, A. (2004). The architecture of complex weighted networks. *Proceedings of the National Academy of Sciences of the United States of America, 101*(11), 3747-3752.

Bhadra, D., & Kee, J. (2008). Structure and dynamics of the core US air travel markets: A basic empirical analysis of domestic passenger demand. *Journal of Air Transport management, 14*(1), 27-39.

Borenstein, S., & Rose, N. (2014). How Airline Markets Work…or Do They? Regulatory Reform in the Airline Industry. In N. Rose, *Economic Regulation and Its Reform: What Have We Learned?* (pp. 63-135). Chicago: University of Chicago Press.

Bureau of Transportation Statistics (2017). 2017 Annual and December U.S. Airline Traffic Data. Retrieved May 4, 2010, from https://www.bts.dot.gov/newsroom/2017-annual-and-december-us-airline-traffic-data

Chen, L., & Miller-Hooks, E. (2012). Resilience: An Indicator of Recovery Capability in Intermodal Freight Transport. *Transportation Science, 46*(1), 109-123.

Chi, L., & Cai, X. (2004). Structural changes caused by error and attack tolerance in US airport network. *International Journal of Modern Physics B, 18*(17-19), 2394-2400.

Cook, G., & Goodwin, J. (2008). Airline Networks: A Comparison of Hub-and-Spoke and Point-to-Point Systems. *Journal of Aviation/Aerospace Education & Research, 17*(2), 51-60.

Crucitti, P., Latora, V., Marchiori, M., & Rapisarda, A. (2003). Efficiency of scale-free networks: error and attack tolerance. *Physica A: Statistical Mechanics and its Applications, 320*, 622-642.

da Rocha, L. (2009). Structural evolution of the Brazilian airport network. *Journal of Statistical Mechanics: Theory and Experiment*.

Dai, L., Derudder, B., & Liu, X., (2018). The evolving structure of the Southeast Asian air transport network through the lens of complex networks, 1979–2012. *Journal of Transport Geography, 68*, 67-77.

Devlin, S., Gnanadesikan, R., & Kettenring, J. (1975). Robust estimation and outlier detection with correlation coefficients. *Biometrika, 62*(3), 531-545.

Dobruszkes, F., & Van Hamme, G. (2011). The impact of the current economic crisis on the geography of air traffic volumes: an empirical analysis. *Journal of Transport Geography, 19*, 1387-1398.

Franke, M. (2004). Competition between network carriers and low-cost carriers—retreat battle or breakthrough to a new level of efficiency? *Journal of Air Transport Management, 10*(1), 15-21.

Freeman, L. C. (1978). Centrality in social networks: conceptual clarification. *Social Networks, 1*, 215-239.

Freeman, L. C., Roeder, D., & Mulholland, R. (1979). Centrality in social networks: ii. experimental results. *Social Networks, 2*(2), 119-141.

Gastner, M., & Newman, M. (2006). Optimal design of spatial distribution networks. *Physical Review E, 74*(1).

Gelhausen, M., Berster, P., & Wilken, D. (2013). Do airport capacity constraints have a serious impact on the future development of air traffic? *Journal of Air Transport Management, 28*, 3-13.

Ghobrial, A., & Irvin, W. (2004). Combating Air Terrorism: Some Implications to the Aviation Industry. *Journal of Air Transportation, 9*(3), 67-86.

Gillen, D., & Morrison, W. (2005). Regulation, competition and network evolution in aviation. *Journal of Air Transport Management, 11*(3), 161-174.

Girvan, M., & Newman, M. (2002). Community structure in social and biological networks. *Proceedings of the National Academy of Sciences, 99*(12), 7821–7826.

Goetz, A., & Vowles, T. (2009). The good, the bad, and the ugly: 30 years of US airline deregulation. *Journal of Transport Geography, 17*(4), 251-263.

Guimerà, R., & Amaral, L. (2004). Modeling the world-wide airport network. *The European Physical Journal B, 38*(2), 381-385.

Guimerà, R., Mossa, S., Turtschi, A., & Amaral, L. (2005). The worldwide air transportation network: Anomalous centrality, community structure, and cities' global roles. *Proceedings of the National Academy of Sciences of the United States of America, 100*(22).

Janić, M. (2015). Modelling the resilience, friability and costs of an air transport network affected by a large-scale disruptive event. *Transportation Research Part A: Policy and Practice, 71*, 1-16.

Jia, T., Qin, K., & Shan, J. (2014). An exploratory analysis on the evolution of the US airport network. *Physica A: Statistical Mechanics and its Applications, 413*, 266-279.

Latora, V., & Marchiori, M. (2001). Efficient Behavior of Small-World Networks. *Physical Review Letters, 87*(19).

Lin, J., & Ban, Y. (2014). The evolving network structure of US airline system during 1990–2010. *Physica A: Statistical Mechanics and its Applications, 410*, 302-312.

Lordan, O., Sallan, J., & Simo, P. (2014a). Study of the topology and robustness of airline route networks from the complex network approach: a survey and research agenda. *Journal of Transport Geography, 37*, 112-120.

Lordan, O., Sallan, J., Simo, P., & Gonzalez-Prieto, D. (2014b). Robustness of the air transport network. *Transportation Research Part E: Logistics and Transportation Review, 68*, 155-163.

Morrison, J., Bonnefoy, P., Hansman, R., & Sgouridis, S. (2010). Investigation of the Impacts of Effective Fuel Cost Increase on the U.S. Air Transportation Network and Fleet. *10th AIAA Aviation Technology, Integration, and Operations (ATIO) Conference.* Fort Worth, Texas.

Pearce, B. (2012). The state of air transport markets and the airline industry after the great recession. *Journal of Air Transport Management, 21*, 3-9.

Redondi, R., Malighetti, P., & Paleari, S. (2012). De-hubbing of airports and their recovery patterns. *Journal of Air Transport Management, 18*(1), 1-4.

Reggiani, A. (2013). Network resilience for transport security: Some methodological considerations. *Transport Policy, 28*, 63-68.

Reynolds-Feighan, A. (1998). The Impact of U.S. Airline Deregulation on Airport Traffic Patterns. *Geographical Analysis, 30*(3), 234-253.

Rocha, L. (2017). Dynamics of air transport networks: A review from a complex systems perspective. *Chinese Journal of Aeronautics, 30*(2), 469-478.

Seidenstat, P. (2004). Terrorism, Airport Security, and the Private Sector. *Review of Policy Research, 21*(3), 275-291.

Sen, A. (1973). *On Economic Inequality.* Oxford: Clarendon Press.

Sun, X., Wandelt, S., & Linke, F. (2015). Temporal evolution analysis of the European air transportation system: air navigation route network and airport network. *Transportmetrica B: Transport Dynamics, 3*(2).

Tam, R., & Hansman, R. (2003). *An Analysis of the Dynamics of the US Commercial Air Transportation System.* SM Thesis, Massachussetts Institute of Technology, MIT International Center for Air Transportation, Cambridge, MA.

Upham, P., Thomas, C., Gillingwater, D., & Raper, D. (2003). Environmental capacity and airport operations: current issues and future prospects. *Journal of Air Transport Management, 9*(3), 145-151.

Vowles, T. (2006). Geographic Perspectives of Air Transportation. *The Professional Geographer, 58*(1), 12-19.

Wandelt, S., & Sun, X. (2015). Evolution of the international air transportation country network from 2002 to 2013. *Transportation Research Part E: Logistics and Transportation Review, 82*, 55-78.
26

Wandelt, S., Sun, X., & Zhang, J. (2017). Evolution of domestic airport networks: a review and comparative analysis. *Transportmetrica B: Transport Dynamics*.

Watts, D., & Strogatz, S. (1998). Collective dynamics of 'small-world' networks. *Nature, 393*, 440-442.

Wilkinson, S., Dunn, S., & Ma, S. (2012). The vulnerability of the European air traffic network to spatial hazards. *Natural Hazards, 60*(3), 1027-1036.

Wuellner, D., Roy, S., & D'Souza, R. (2010). Resilience and rewiring of the passenger airline networks in the United States. *Physical Review E, 82*(5).

Xu, Z., & Harriss, R. (2008). Exploring the structure of the U.S. intercity passenger air transportation network: a weighted complex network approach. *GeoJournal, 73*, 87-102.

Zhang, J., Cao, X., Du, W., & Cai, K. (2010). Evolution of Chinese airport network. *Physica A: Statistical Mechanics and its Applications, 389*(18), 3922-3931.

Zhou, Y., Wang., J., & Huang., G. Q. (2019). Eficiency and robustness of weighted air transport networks. *Transportation Research Part E, 122*, 14-26.